\documentclass[twocolumn,preprintnumbers,amsmath,amssymb,showpacs]{revtex4-1}
\usepackage{graphicx}
\usepackage{dcolumn}
\usepackage{bm}
\usepackage{lipsum}

\begin{document}


\title{Hide the interior region of core-shell nanoparticles with quantum invisible cloaks}
\author{Jeng Yi Lee$^{1}$ and Ray-Kuang Lee$^{1,2}$}

\affiliation{
$^1$ Institute of Photonics Technologies, National Tsing-Hua University, Hsinchu 300, Taiwan\\
$^2$ Frontier Research Center on Fundamental and Applied Sciences of Matters, National Tsing-Hua University, Hsinchu 300, Taiwan}
\date{\today}

\begin{abstract}
By applying the interplay among the nodal points of partial waves, along with the concept of streamline in fluid dynamics for the probability flux,  a quantum invisible cloak to the  electron transport in a host semiconductor is demonstrated by simultaneously guiding the probability flux outside the core region and keeping the total scattering cross section negligible.
As the probability flux vanishes in the interior region,  one can embed any material inside  a multiple core-shell sphere without affecting physical observables from the outside.
Our results reveal the  possibility to  design a protection shield layer for  fragile interior parts  from the impact of transports of  electrons.
\end{abstract}

\maketitle
Transformation method, originally proposed to control electromagnetic (EM) waves ~\cite{Science1, Science2},  has been applied to a broad area of physics from optics, acoustics, to  thermo-electricity. 
By avoiding any scattering and preventing waves from entering the interior region, {\it invisible cloak} devices are realized experimentally~\cite{exp}.
As a member of wave family, search  for quantum transformation thoery for  matter waves is proposed theoretically, but limited by the required spatial distribution of complicated effective mass and potential~\cite{1, 2, 3, 5}.
Instead of using  functionally graded meta-materials, another possible way to have cloaking devices is achieved by scattering cancellation.
By eliminating the leading terms in scattering cross sections,  a small scatter,  with the size compared to the  light wavelength,  coated by isotropic and homogeneous meta-material or surface plasmons can be used to greatly reduce the EM scattering waves~\cite{9, 10}.
Consequently, due to the similar mathematical structure between Schr\"odinger equation and Maxwell equation, a core-shell nanoparticle  embedded in a host semiconductor is found to become invisible for the matter wave, as the de Broglie wavelength of transport electrons can be compatible or longer than the scatter size~\cite{8, 7}. 
Extension of this approach is also applied to enhance power efficiency in thermoelectric devices by using invisible dopants~\cite{11}. 

\begin{figure}[b]
\centering
\includegraphics[width=8.4cm]{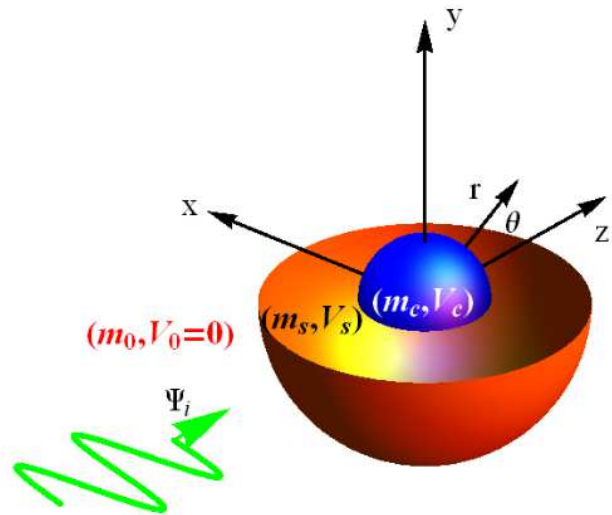}
\caption{\label{fig:fig1}(Color online) A core-shell nanoparticle with different effective masses and potential energies defined in each region.  Quantum matter wave of the transport electron, $\Psi_i$, is assumed to propagate along the $z$-axis.}
\end{figure}

In a barrier-well potential, it is the {\it push-and-pull} interference between the two dominant lowest-order partial waves that makes the total scattering cross section negligible~\cite{7}.
Even though the core-shell nanoparticle is invisible from not being  detected by the conduction electrons with a typical energy scale in semiconductors,  the probability flux of matter wave still passes through the entire nanoparticle.
In this scenario,  unless the required cloaking parameters of a delicate functional device matches well, a slight change of physical parameters in the core  may result in a severe degradation on the scattering cross section ~\cite{8}. 
As transformation optics demonstrates, an {\it invisible cloak} should  prevent fields from entering the region, as well as avoid any scattering in the region around simultaneously.
In order to hide from the electron transport, to be invisible and cloaking at the same time,
in this Letter,  we demonstrate the possibility on making  nanoparticles invisible as well as guiding the probability flux outside the interior region, by applying the concept of streamline in fluid dynamics.
To guide probability flux of quantum particle waves outside  the cloaking region, a destructive interference between incoming and outgoing spherical traveling waves in the shell region is required, resulting in the total internal reflection at the core-shell interface.
Furthermore, by considering nanoparticles in shape of a multiple core-shell sphere,  we reveal an invisible cloak for the  transport of conducting electrons, in which one can embed any material inside the interior region without affecting physical observables from the outside. 
Based on these results, we provide the guideline to design a device to protect fragile interior parts  from the impact of large conduction electrons. 

As illustrated in Fig. 1, we begin with a spherical nanoparticle,  composed of two concentric layers of homogeneous isotropic materials, with different effective masses and potential energies in each region.
Consider a conduction electron with the energy $E$, an effective mass $m_{0}$, and  the characteristic incident plane matter wave $\Psi_{i}$ propagating along the $z$-axis.
The corresponding wavenumber outside the nanoparticle is defined as  $k_0 = \sqrt{2m_{0}(E-V_0)}/\hbar$, where the potential energy  in the background  environment is set to zeo,  $V_0=0$. 
In the shell region, the effective mass and potential are denoted by $m_{s}$ and  $V_{s}$, with the corresponding wavenumber $k_{s}=\sqrt{2m_{s}(E-V_{s})}/\hbar$; while in the core region,   the effective mass, potential, and wavenumber are denoted as $m_{c}$, $V_{c}$,  and $k_{c}=\sqrt{2m_{c}(E-V_{c})}/\hbar$, respectively.
The potential energies are chosen that $V_s < 0$ and $V_c > 0$ for  a core-shell nanoparticle.
Geometrical parameters for the core-shell nanoparticles are the radius of core, $a_{c}$, and the radius of whole particle, $a$. 

Without loss of generality, we  limit ourself to the case $k_0 a \le  1$  when the size of nano-particle is the smallest length scale.
Within the typical energy scale,  conduction electrons transport through nanoparticles in a host semiconductor would have a sufficiently long de Broglie wavelength, as pointed out by Liao {\it et al.} \cite{7}.
Following the partial wave analysis for Schr\"odinger equation with an incident plane wave $\Psi_{i}$, {\it i.e.}, our method is also applied to the Mie's scattering theory for EM waves ~\cite{book-Gbur}, one can decompose the matter wave solution outside the nanoparticle into $\Psi_0(r,\theta) =\Psi_{i}+\Psi_{scat} =\sum_{l=0}^{l=\infty}i^{l}(2l+1)\{j_{l}(k_0 r)+a_{l}^{scat}h^{(1)}_{l}(k_0 r)\}P_{l}(\cos\theta)$, with the complex coefficient  $a_{l}^{scat}$ for scattering matter wave $\Psi_{scat}$. 
By the same way, the corresponding wave solutions in the shell and core regions have the form: 
$\Psi_{s}(r,\theta) = \sum_{l=0}^{l=\infty}i^{l}(2l+1)\lbrace  b_{l}j_{l}(k_{s}r)+c_{l}h^{
(1)}_{l}(k_{s}r)\rbrace P_{l}(\cos\theta)$ and $\Psi_{c}(r,\theta)=\sum_{l=0}^{l=\infty}i^{l}(2l+1)d_{l}j_{l}(k_{c}r) P_{l}(\cos\theta)$, respectively.
Here, $b_{l}$, $c_{l}$, and $d_{l}$ are the characteristic complex coefficients for the infinite series of partial wave in each region, with  $j_{l}$, $h^{(1)}_{l}$, and $P_{l}$ being the $l$th-order spherical Bessel function, spherical Hankel function of the first kind, and Legendre function, respectively. 
By matching the boundary conditions at each interface to satisfy the continuity of wave function and conservation of probability flux in the radial direction, $r$, we can find the corresponding complex coefficients.
 
For a given transport electron energy $E$ and a fixed geometric parameter set for core-shell nanoparticles, {\it i.e.}, the radii $a$ and $a_{c}$, it is possible to find  a  range of parameters to support a negligible total scattering cross section~\cite{8,7}.
Mathematically, when the condition $k_0 a \le 1$ in each region is satisfied, one can only consider the first two lowest-order terms as the  dominant ones in the scattering cross section expansion, {\it i.e.},  $l = 0$ and $l = 1$, coined as the $s$-waves  and $p$-waves, respectively. 
In this case, to have scattering cancellation from the $s$-waves and $p$-waves,  a phase difference of $\pi$ is implemented by varying the  corresponding effective masses and potential energies accordingly, in order to induce a destructive interference.
However, we want to point out that only two degrees of freedom in the parameter set, $\{m_{s},  V_{s}, m_{c}, V_{c}\}$, are used in this quantum cloaking technique~\cite{8,7}.
Below,  we show that the other two degrees of freedom can be used to satisfy
 a zero probability flux inside the interior region by requiring a nodal point solution around the core-shell interface for the wave solution in the shell region.

\begin{figure}[t]
\centering
\includegraphics[width=8.4cm]{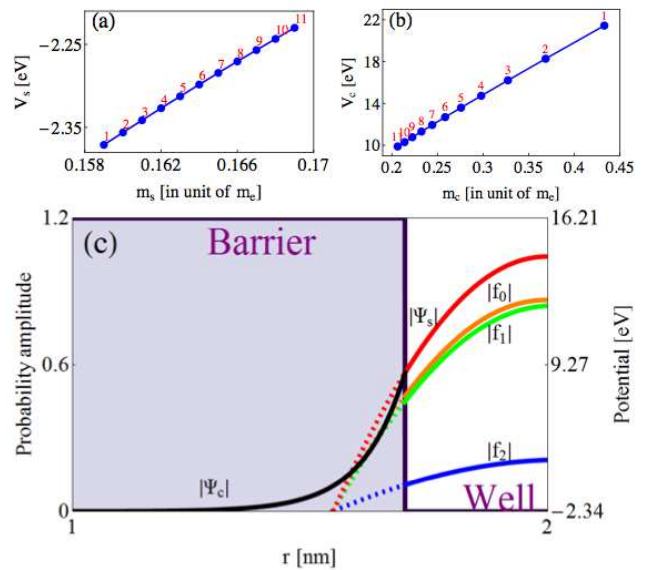}
\caption{\label{fig:fig2}(Color online) (a) The parameter set $\{m_s, V_s\}$ to support a total internal reflection for the wave function in the shell region, $\Psi_s$; and (b) the corresponding parameter set $\{m_c, V_c\}$, marked in numbers,  to simultaneously generate a negligible total scattering cross section within $10^{-4}$.  (c) The probability amplitude of wave functions inside the barrier-well potential, as a function of the radial axis $r$, are shown for $|\Psi_c|$ and $|\Psi_s|$ in black and red colors, respectively. Decomposed channel functions $f_l$, $l = 0, 1, 2$, are also shown as a comparison. Moreover, dashed-lines in the barrier region show the extension of wave functions from the shell (potential well) region, all of which have a common nodal point. 
The parameter set  $\{m_s, V_s, m_c, V_c\}$ used in (c) are $\{0.16m_e$, $-2.34$eV, $0.33m_e$, $16.21$eV$\}$, with the geometric size parameters of nanoparticle $a = 2$nm and $a_{c} = 1.7$nm. The energy and effective mass for the transport electron are $E = 0.01$eV  and $m_{0}=0.8 m_{e}$ (in unit of the electron mass $m_e$), respectively.}
\end{figure}

To illustrate our method, first of all, we assume that there exists a set of parameters to support an invisible nanoparticle (this assumption should be verified afterward).
In this case, we have $a_{l}^{scat}=0$ for $l=0,1$ and $a_{l}^{scat}\approx 0$  for $l \ge 2$ (its absolute value is smaller than $10^{-4}$). 
Then, under above approximation, the corresponding coefficients  $b_{l}$ and $c_{l}$ for the wave function in the shell region, $\Psi_s$, can be found in a closed form,
\begin{eqnarray}
b_{l}&\approx &b_{l}^{app}=\frac{x_{2}y_{1}j_{l}(x_{1})h_{l}^{(1)'}(x_{2})-x_{1}y_{2}h_{l}^{(1)}(x_{2})j_{l}^{'}(x_{1})}{x_{2}y_{1}j_{l}
(x_{2})h_{l}^{(1)'}(x_{2})-x_{2}y_{1}h^{(1)}_{l}(x_{2})j^{'}_{l}(x_{2})},\nonumber\\
\\
c_{l}&\approx &c_{l}^{app}=\frac{x_{1}y_{2}j_{l}(x_{2})j_{l}^{'}(x_{1})-x_{2}y_{1}j_{l}(x_{1})j^{'}_{l}(x_{2})}{x_{2}y_{1}j_{l}(x_{2})h_{l}^{(1)'}(x_{2})-x_{2}y_{1}h^{(1)}_{l}(x_{2})j^{'}_{l}(x_{2})},\nonumber \\
\end{eqnarray}
where the shorthanded notations used are $j^{'}_{l}(x)\equiv{\text{d}j_{l}(x)}/{\text{d}x}$, $h_{l}^{(1)'}(x)\equiv {\text{d}h_{l}^{(1)}(x)}/{\text{d}x}$, $x_{1}\equiv k_0 a$, $y_{1}\equiv m_{0}a$, $x_{2}\equiv k_{s}a$, and $y_{2}\equiv m_{s}a$.
Physically, coefficients shown in Eqs. (1-2) represent  a set of solutions for the outgoing and incoming spherical traveling waves in the shell region.
By transforming these solution bases into the pair of Hankel functions of the first and the second kinds, $h_{l}^{(1)}$ and $h_{l}^{(2)}$, the corresponding coefficients for the outgoing and incoming waves are $i^{l}(2l+1)(b^{app}_{l}/2+c^{app}_{l})$ and ${i^{l}}(2l+1)b^{app}_{l}/2	$, respectively ~\cite{13}.
Moreover,  the condition $|b^{app}_{l}/2+c^{app}_{l}|=|b^{app}_{l}/2|$ is valid for any real parameters, which means that each partial wave forms a kind of standing waves in the shell region.
With the above physical insights, we look for a total internal reflection at the core-shell interface, by imposing a destructive interference between the outgoing and incoming spherical traveling waves. 
As the Goos-H\"anchen phase shift happens in the total internal reflection ~\cite{Haus},  here, we look for the wave solution $\Psi_{s}$ with a nodal point outside the shell region, in
 order to create a cloaking interior region.
In other words, instead of finding the separating nodal point for each partial wave inside the shell region, our first condition is to set the nodal points of $\Psi_s$ penetrating a little into the core region.

In addition, by the conservation of probability flux, one has $\vec{\triangledown}\cdot\vec{J}_i=0$, where $\vec{J}_i={e\hbar}/{m_{i}}\,\text{Im}[\Psi_{i}^{*}\triangledown\Psi_{i}]$, with  $m_{i}$ and $\Psi_{i}$ corresponding to the effective mass and wave function in each region.
To have an invisible cloak, we assume that the probability flux in the shell region,  $\vec{J}_{s}$, is approximately orthogonal to $x$-$y$ plane at the polar angle $\theta=\frac{\pi}{2}$, that is
\begin{equation}\label{conservation}
\int_0^{2\pi}\int_{a_{c}}^{a}\hat{z}\cdot\vec{J}_{s}(r,\theta=\frac{1}{2}\pi)r \text{d}r\text{d}\phi= \pi a^{2}\frac{e\hbar k_0}{m_{0}}(1-\epsilon),
\end{equation}
with a small free parameter $\epsilon$ used to  measure the loss of total flux due to any  possible quantum tunneling  effect occurred at the potential interfaces.
In general, $\epsilon$ can be an infinitesimal value, but to have a practical parameter set, in the following, we set $\epsilon=0.05$ as an illustration.
Basically, as the case in fluid dynamics, Eq. (3) enforces the streamline of  probability flux to flow only in the shell region~\cite{fluid}.

\begin{figure}
\centering
\includegraphics[width=8.4cm]{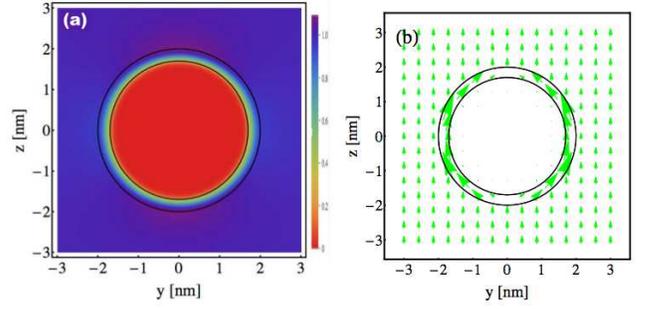}
\caption{\label{fig:fig3}(Color online) (a) Probability amplitude $|\Psi|$, and (b) the corresponding probability flux $\vec{J}$ of the wave function in the plane $x = 0$ for a quantum invisible cloak, with the same parameter set used in Fig. 2(c).
In this simulation, the probability inside the core region is below $10^{-10}$; while near the interface of core-shell, a very high flux is generated  to satisfy the conservation of probability flux.}
\end{figure}

With the introduction of these two additional conditions: to seek a nodal point for the wave solution $\Psi_s$ penetrating into the core region due to the total internal reflection,  and to have a  conserved probability flux in the shell region, it is sufficient to determine two physical parameters in the shell region.
In Fig. 2(a), we report the parameter set  $\{m_{s}, V_{s}\}$ to support a quantum invisible cloak numerically  for the wave function in the shell region $\Psi_s$ with a nodal point within the core region, $|\Psi_s(r_n)| = 0$ at $r_n < a_c$. 
The found wave function in the shell region is shown in Fig. 2(c), with the comparison to the decomposed channel function defined as $f_{l}(r,\theta = 0)=i^{l}(2l+1)\lbrace  b_{l}j_{l}(k_{s}r)+c_{l}h^{(1)}_{l}(k_{s}r)\rbrace P_{l}(\cos\theta)$, for different $l$.
As shown in Fig. 2(c), the contribution from higher-order terms of partial waves are significantly small enough as expected. 
Moreover, such a nodal point, penetrating into the core region, exists as the case of  Goos-H\"anchen phase shift in the total internal reflection.
Since we have a spherical symmetry in geometry, the requirement to have a nodal point solution for all the polar angle $\theta$ results in guiding the probability flux to  flow only in the shell region. The non-trivial solution to satisfy $\vec{\triangledown}\cdot\vec{J}_i=0$ demands the streamline of probability flux to circulate around the core region, as shown in Fig. 3(b).

Then, with the set of solutions $\{m_{s}, V_{s}\}$ found in Fig. 2(a), we search for the corresponding set of parameters  $\{m_{c}, V_{c}\}$ by coming back to the concept of scattering cancellation ~\cite{8, 7}.
The scattering cross section outside the nanoparticle has the form,
\begin{equation}\label{scattering}
\sigma=\frac{4\pi}{k_0^{2}}\sum_{l=0}^{l=\infty}(2l+1)|a_{l}^{scat}|^{2},
\end{equation}
with the complex scattering coefficient $a_l^{scat}$ defined as
\begin{equation}
a_{l}^{scat}=-\frac{x_{1}j^{'}_{l}(x_{1})[A_{l}+B_{l}]+y_{1}j_{l}(x_{1})[C_{l}+D_{l}]}{x_{1}h^{'}_{l}(x_{1})[A_{l}+B_{l}]+y_{1}h_{l}(x_{1})[C_{l}+D_{l}]},
\end{equation}
where
\begin{eqnarray}
A_{l}=y_{2}x_{3}y_{4}j_{l}(x_{4})[j_{l}(x_{2})h^{'}_{l}(x_{3})-h_{l}(x_{2})j_{l}^{'}(x_{3})],\\
B_{l}=y_{2}y_{3}x_{4}j_{l}^{'}(x_{4})[h_{l}(x_{2})j_{l}(x_{3})-j_{l}(x_{2})h_{l}(x_{3})],\\
C_{l}=x_{2}x_{3}y_{4}j_{l}(x_{4})[h_{l}^{'}(x_{2})j_{l}^{'}(x_{3})-h_{l}^{'}(x_{3})j_{l}^{'}(x_{2})],\\
D_{l}=x_{2}y_{3}x_{4}j_{l}^{'}(x_{4})[j_{l}^{'}(x_{2})h_{l}(x_{3})-h^{'}_{l}(x_{2})j_{l}(x_{3})].
\end{eqnarray}
Here, the shorthanded notations used are $x_{3}\equiv k_{s}a_{c}$, $x_{4}\equiv k_{c}a_{c}$, $y_{3}\equiv m_{s}a_{c}$, and $y_{4}\equiv m_{c}a_{c}$. 
To have a  negligible scattering cross section, Alu {\it et al.} has shown that $k_0 a$, $k_s a$, and $k_c a$ should be all smaller than $1$, then only the components of $s$- and $p$-waves come to play a dominant role. 
However, in our case to support a negligible scattering cross section, we have $|k_s a| = 6.30$ and $|k_c a| = 20.06$, which are both greater than $1$.
The set of parameters $\{m_c, V_c\}$ found to support an invisible cloak is shown in Fig. 2(b), with the markers corresponds to the set of parameters $\{m_s, V_s\}$ shown in Fig. 2(a) for confining the probability flux in the shell region simultaneously.
By our proposed method, the parameter set $\{m_s, V_s, m_c, V_c\}$ generates a quantum invisible cloak, which has almost zero values both for the total scattering cross section and  probability flux in the core region simultaneously.
In Fig. 3, we plot the distributions of probability amplitude and probability  flux for a quantum invisible cloak, with the the parameters $a=2$nm, $a_{c}=1.7$nm, $m_{0}=0.8m_{e}$, $E=0.01$eV, $m_s = 0.16m_e$, $V_s = -2.34$eV, $m_c = 0.33m_e$, and $V_c = 16.21$eV, respectively.
It can be seen that almost all the probability flux circulates along the boundary of barrier-well potential, while inside the core region the probability flux is almost zero, as shown in Fig. 3(a) and 3(b).
Take the demonstration shown in Fig. 3 as an example, the probability to find matter wave inside the core region is as small as $10^{-10}$ at $r = 0.5 a_c$, and about $10^{-15}$ at $r=0$, along with keeping the total scattering cross section at $10^{-4}$.
We want  to remark that in the calculations, we do not restrict ourself by truncating the infinite series for the scattering cross section and  the corresponding probability flux.
Instead,  we calculate as many as possible terms to satisfy the  conditions that the value of each terms is smaller than $10^{-5}$.

\begin{figure}
\includegraphics[width=8.4cm]{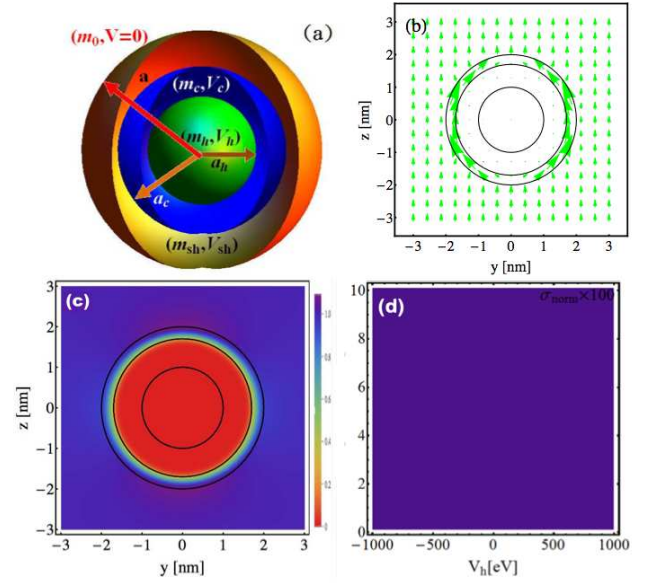}
\caption{\label{fig:fig4}(Color online) (a) Illustration of a multiple core-shell nanoparticle with $3$ layers in the structure. The effective masses and potential energies in each region are denoted by $\{m_s, V_s\}$, $\{m_c, V_c\}$, and $\{m_h, V_h\}$, with the corresponding radii $a$, $a_c$, and $a_h$, respectively.  Probability flux and the corresponding probability amplitude  for such a quantum invisible cloak are shown in (b) and (c) with the parameters: $a=2$nm, $a_{c}=1.7$nm, $a_h = 1$nm, $m_{0}= 0.8 m_{e}$, $m_{s}=0.16m_{e}$, $m_{c}=0.33 m_{e}$, $m_h = 0.055 me$, $V_{s}=-2.34$eV, $V_{c}=16.21$eV, $V_h = -9000$eV, and $E=0.01$eV, respectively.
In (d), for a wide range of effective mass inside the hidden region, $m_h = [0, 10 m_e]$,  the corresponding scattering cross section remains almost unchanged no matter how an extremely large positive or negative value of the potential energy $V_h$ is.
In all of these cases, the normalized scattering cross section is $10^{-4}$.} 
\end{figure}

However, since the nodal point of wave function is chosen to be located outside the sell region, a little amount of the probability flux, about $\epsilon\pi a^2 e\hbar k_0/m_0$, penetrates in the core region.
In this case,  for a single barrier-well potential only,  we can  not  take this core region as an ideal hidden region. 
To hide the interior region of a nanoparticle, we extend our method for a nanoparticle in shape of multiple core-shell spheres.
Take a three-layer potential as an example shown in Fig. 4(a), now, it is possible to have a hidden interior region for $r < a_h$, where $a_h$ is the radius of the interior hidden region.
If the secondary layer is thin, as expected, a slight change of any parameter  in the hidden region would strongly affect the scattering properties in the outside  (not shown here).
Instead, if the secondary layer is thick enough to exclude the quantum tunneling effect, then the total scattering cross section remains unchanged no matter what kind of the material is placed inside the interior region, as shown in Fig. 4(d).
This result implies that our hidden region in a three-layer structure is indeed strongly robust even for this extremely set of parameters.  
Range of potential parameters to  realize such a  quantum invisible cloak may also be provided with the present semiconductor technologies.

Before conclusion, we try to discuss the difference in one-dimensional (1D) and two-dimensional (2D) systems. 
Consider an unbound state (scattering state) in 1D system, transmission resonance may happen when a free electron encounters a rectangular barrier.
In this case, the front scattering wave and the back one have a chance to interfere and cancel each other, but the wave needs to penetrate the entire region ~\cite{14}.
As a result, one can not have a shield region in 1D.
In the other way, for a bound state in 1D, one may create a shield region by using two rectangular potential barriers with a suitable separation. 
Then, these two isolated wave functions can exist in each region by separating the barriers, but there is no net probability flux to flow, in contrast to the  transmission resonance. 
However, even only with the radial Schr\"odinger equation  considered here, there is an extra degree of dimensions to guide the probability flux.
Then, with suitable parameters, one can have the probability flux to  flow only in the shell region, without entering the core.
Therefore, we can provide the transmission resonance (invisible) and shield (cloaking) simultaneously. 
This is a completely different situation compared to that in one dimension, as the application of conformal mapping happens in two dimensions.
Moreover, while the total scattering cross section is kept negligible, the condition to require $k_c a$ and $k_s a$ both smaller than $1$ is also relaxed in our study, which differs considerably from earlier studies~\cite{8, 10}.

In summary, we introduce two more degrees of freedom in designing a quantum invisible cloak for core-shell nanoparticles by requiring the total internal reflection and conservation of the probability flux in the shell region.
Solutions of wave function in the shell region  need to have a nodal point penetrating into the core region,  resulting in  guiding the probability flux outside the core region as  streamline does in fluid dynamics.
In such a way, we can embed any physical material inside the hidden region without disturbing the environmental probability. 
Moreover, this hidden region can also provide a protective shield for some sensitive or easily corrupted devices from the impact of transports of electrons. 
With the analogy between quantum matter waves and classical waves, the concept of our method can be ready applied in other fields, such as electromagnetic and acoustic systems, etc.

\end{document}